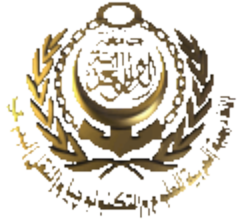

Arab Academy for Science and Technology and Maritime Transport

(AASTMT)

College of Engineering and Technology

Department of Computer Engineering

# TOWARDS METAMORPHIC VIRUS RECOGNITION USING EIGENVIRUSES

A dissertation submitted in partial fulfillment of the requirements for the degree of Master of Science in Computer Engineering

By

## Moustafa Saleh

B.Sc. Comp. Eng. Arab Academy for Science and Technology and Maritime Transport

Supervisors

## Prof. Dr. A. Baith Mohamed

Professor, Computer Engineering
College of Engineering & Technology
(AASTMT)

## Dr. Ahmed A. Nabi

Associate Professor, Head of Network and distributed systems department, Informatics Research Institute, Mubarak City for scientific Research and Technology Applications

July 2011

# Declaration

The work in this dissertation is based on researches that were carried out at the Department of Computer Engineering, Arab Academy for Science and Technology, Alexandria, Egypt. No part of this dissertation has been submitted elsewhere for any other degree or qualification and it is all my own work unless referenced to the contrary in the text. The contents of this dissertation reflect my own personal views, and are not necessarily endorsed by the University.

Signature:                                    Date:

# Declaration

We certify that we have read the present work and that in our opinion it is fully adequate in scope and quality as a dissertation towards the partial fulfillment of the Master degree requirements in Computer Engineering from College of Engineering and Technology, Arab Academy for Science and Technology and Maritime Transport.

Supervisors:

| | |
|---|---|
| Name: | |
| Position: | |
| Signature: | |
| Name: | |
| Position: | |
| Signature: | |

Examiners:

| | |
|---|---|
| Name: | |
| Position: | |
| Signature: | |
| Name: | |
| Position: | |
| Signature: | |

*Dedication*

*For the children who lost their homes and parents in Gaza war*

*And still hungry under blockade,*

*Never give up,*

*Never let down*

# Acknowledgments

First of all, thanks go to my God, then, I would like to thank my parents. My mother, who has always been there for me and taught me to be a hard worker, willing to learn and never lose hope. My father who instilled in me the love of reading from the moment he began bringing me books when I was five years old.

I would like to thank Dr. Tahir El-Sunni for introducing me to the world of multivariate statistics by assigning me the presentation of Principal Component Analysis. I would also like to give special thanks to Dr. Sherin Youssef who has a smart insight of her students that made her push me to choose the topic of Eigenfaces in Artificial Intelligence course. Without comprehension of Principal Component Analysis and Eigenfaces, I could not come up with this thesis. I also take advantage of this to thank all my instructors in my educational life.

Big thanks to Peter Szor, the chief antivirus researcher at Symantec Corporation and Peter Ferrie, senior antivirus researcher at Microsoft Corporation for their cooperation and responds to my questions regarding techniques used by commercial antivirus software and effectiveness of published experimental detection methods.

Special thanks go to Thomas Sperl (aka SPTH), the author of W32/Flibi worm. His help and excitement made me rediscover the strong potentials of my thesis.

I owe a huge thanks to my brother Ahmed Saleh, who has been always helping me out in all my life. He was always the best brother and friend for me.

Last but not least, I would like to thank my wife who has been always patient and supportive to me when I spent many nights and weekends away working on this thesis.

# Abstract

Metamorphic viruses are considered the most dangerous of all computer viruses. Unlike other computer viruses that can be detected statically using static signature technique or dynamically using emulators, metamorphic viruses change their code to avoid such detection techniques. This makes metamorphic viruses a real challenge for computer security researchers. In this thesis, we investigate the techniques used by metamorphic viruses to alter their code, such as trivial code insertion, instructions substitution, subroutines permutation and register renaming. An in-depth survey of the current techniques used for detection of this kind of viruses is presented. We discuss techniques that are used by commercial antivirus products, and those introduced in scientific researches.

Moreover, a novel approach is then introduced for metamorphic virus recognition based on unsupervised machine learning generally and Eigenfaces technique specifically which is widely used for face recognition. We analyze the performance of the proposed technique and show the experimental results compared to results of well-known antivirus engines. Finally, we discuss the future and potential enhancements of the proposed approach to detect more and other target viruses.

# Table of Contents

# List of Figures

# List of Tables

# List of Symbols

| | |
|---|---|
| **API** | Application Programming Interface |
| **AV** | Anti-Virus |
| **Bistro.B** | Second Version of Bistro Windows Virus |
| **DFA** | Deterministic Finite Automata |
| **DNA** | Deoxyribonucleic Acid |
| **HMM** | Hidden Markov Model |
| **IA-32** | Inter Architecture, 32-bit |
| **IMM** | Immediate Operand |
| **IMM16** | 16-Bit Immediate Operand |
| **IMM32** | 32-Bit Immediate Operand |
| **In-The-Wild** | Active and spreading virus |
| **MEM** | Memory Address |
| **MEM16** | 16-Bit Memory Address |
| **MEM32** | 32-Bit Memory Address |
| **MetaPHOR** | Highly metamorphic virus, also known as Etap and Simile |
| **MOV** | "Move" Instruction |
| **NOP** | "No Operation" Instruction |
| **PCA** | Principal Component Analysis |
| **PE** | Portable Executable |
| **REG** | General Purpose Register |
| **REG8** | 8-Bit General Purpose Register |
| **REG16** | 16-Bit General Purpose Register |
| **REG32** | 32-Bit General Purpose Register |
| **RegEx** | Regular Expressions |
| **SHL** | "Shift Left" Instruction |
| **SVM** | Support Vector Machine |
| **tRNA** | Transfer Ribonucleic Acid |
| **Whale** | Windows virus |
| **Win32** | Family of Windows 32-bit operating systems |
| **Win95** | Windows 95 operating system |
| **Zmist** | Highly metamorphic virus |
| **Zperm** | Highly metamorphic virus |

# Nomenclature

| | |
|---|---|
| $\boldsymbol{\Phi}$ | Column vector represents a virus replicates |
| **A** | N x M matrix that holds all virus replicates |
| **M** | Total number of training files |
| **N** | Eigenvirus length, i.e. Maximum Virus size in bytes |
| **C** | Covariance matrix of **A** |
| *u* | Eigenvectors of the covariance matrix C |
| λ | Eigenvalues of A |
| v | Eigenvectors of A |
| **M`** | Chosen set of eigenvectors with higher eigenvalues |
| $\omega_j$ | Weight of a virus on dimension *j* |
| $\Omega_i$ | A vector contains the weights of virus *i* on each dimension |
| ε | Virus class distance |
| θ | Virus class distance threshold |
| α | Virus space distance |
| β | Virus space distance threshold |

# List of Publications

- A. Abdel Nabi, M. E. Saleh, A. Baith Mohamed, "Eigenviruses for Metamorphic Virus Recognition", *IET journal of information security*, to be published, 2011.

# CHAPTER: I
# INTRODUCTION

# 1  INTRODUCTION

COMPUTER virus is a self-replicating piece of code that attaches itself to other programs and usually requires human interaction to propagate [1]. Computer virus is one of many types of malware that are intentionally created to harm computer systems. Professionally speaking, malware is short for malicious software and it is a general term used to describe any software that is harmful to any scale of computer systems [2][3]. Computer malware analysis and detection is considered a critical topic in computer security, not only because of the significant wide spread of malware, but also because of its economical impact [4].

Every year thousands of new malware arise and cost the world billions of dollars. A survey conducted by "ComputerEconomics.com" indicated that malware economical impact in 2006 exceeded 13 billion Dollars [5]. Another survey shows that consumers in United States only paid about 7.8 billion Dollars over two years to repair or replace computer systems affected by malware [6]. United States Government Accountability Office stated that in the year 2005 the American economy lost about 67.2 billion Dollars because of cybercrimes [7]. Unfortunately, Malware techniques are becoming increasingly sophisticated and the number of new malware doubles each year than the year before. According to F-Secure antivirus Corporation, there were as much malware produced in 2007 as in the previous 20 years altogether [8]. Another disturbing fact is that in the year 2007 the release rate of malicious code and other unwanted programs may be exceeding that of legitimate software applications [9]. Therefore, security researchers and professionals have to find more powerful and effective solutions to keep up with the explosive growth of malware.

One of the oldest types of malware is the computer virus. The term was explicitly mentioned in Fred Cohen researches that theoretically formalized the problem of self-replicating software that he called computer virus. Fred Cohen's definition of computer virus is that it is "a program that can 'infect' other programs by modifying them to include a possibly evolved copy of itself" [10].

Since the first virus appeared in the wild, not only millions of different viruses emerged and attacked computers, but also computer viruses have evolved a lot a long past decades. The latest type of computer viruses is the metamorphic virus. Metamorphic virus is a type of viruses that changes its appearance constantly on each infection, yet maintains the same behavior. Because of this change in appearance, common simple detection techniques such as string signature scanning are useless against metamorphic viruses [11].

## 1.1 Motivation

Several techniques have been proposed to find a solution for metamorphic virus detection. However, many of them were unable to reach commercial products due to their unacceptable rate of false-positive errors or their high computational complexity. Few other methods are widely used, though existing of some limitations and false-positive errors motivate the move to find a more reliable and robust way to detect metamorphic viruses.

In this thesis, we try to tackle the problem of metamorphic virus detection by introducing a novel approach based on Eigenfaces technique that is used in face recognition problem and use it to detect metamorphic viruses.

## 1.2 Contribution

The aim of this thesis is to develop a method to detect metamorphic code with least false-positives errors. In our experiment, we tested five well-known metamorphic viruses against the system. With very small number of training samples of each virus, we succeeded to detect 250 samples of each virus. That is, a 100% successful detection rate was achieved. A set of benign files taken from Cygwin tools package we tested against the system to measure the false-positive error rate. The system was able to identify 244 as clean files, that is, 2.4% false-positive error rate resulted.

## 1.3 Thesis Outline

This thesis is organized into four chapters:

- Chapter One – Introduction: Gives a brief overview about the problem, motivation and the contribution presented in this thesis.
- Chapter Two – Background: It discusses the various stages computer viruses have been through to reach the metamorphic type. In addition, the chapter discusses the different methods used to achieve code metamorphism.

- <u>Chapter Three – Eigenviruses</u>: The chapter introduces eigenviruses technique and gives the mathematical background behind it. Also, it explains the experiment that is carried out to measure the performance of the technique. In the last section of the chapter, after a detailed analysis of the experiment results, the section lists the results of a random copy of each virus used in the experiment when tested against some commercial antivirus products.

- <u>Chapter Four – Conclusion and Future Work</u>: This chapter summarizes the thesis and gives some possible ideas to enhance and extend the proposed system.

# CHAPTER: II
# BACKGROUND

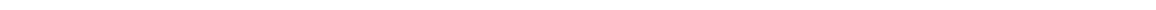

# 2 BACKGROUND

Computer viruses have evolved a lot in the past years. Whenever a new technique is used to countermeasure them, virus writers found a new way to make their viruses harder and more complex. This chapter discusses the stages of evolution the computer viruses have been through until the appearances of metamorphic type of viruses. Also an overview of the metamorphism techniques used by this type of viruses is given at the end of the chapter.

## 2.1 Virus Evolution

We can basically divide the evolution stages of computer viruses into four stages as shown in Figure 2.1. Plain viruses were the first generation of computer viruses. When a plain virus infects a host file, it simply copies itself as it is, thus maintains the same structure along its generations. They were executed exactly as they were written each time they run. Soon, virus researchers could distinguish each virus with a unique pattern of bytes that resembles its signature. Therefore, those viruses can be easily detected by their signature.

Signature detection is very effective in virus detection, in which antivirus software searches for unique constant pattern or sequence of bytes in the virus body [11]. For example, the signature of the virus segment in Figure 2.2 is:

**BE04000000 8BDD B9 0C000000 81C088000000 8B38 89BC8B18110000 2BC6 49**

Consequently, virus writers had to evolve their code in order to evade detection,

and so self-encrypting viruses emerged.

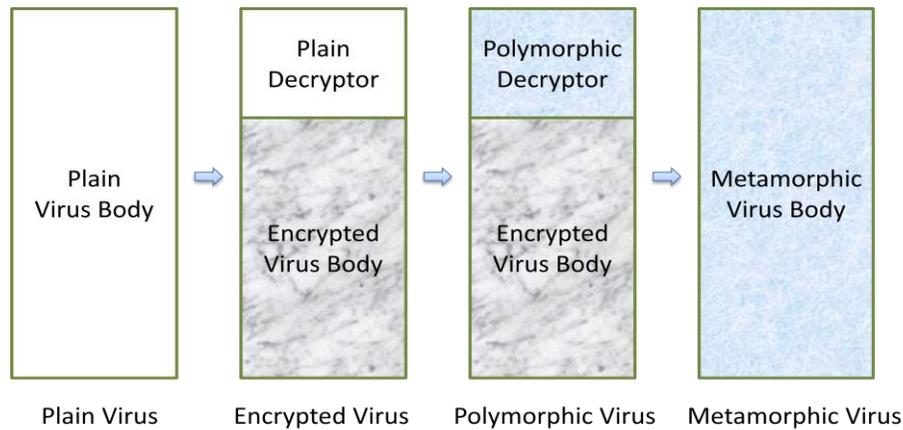

Figure 2.1  Virus Evolution

```
a)
BE04000000        mov     esi,000000004 ;" ?"
8BDD              mov     ebx,ebp
B90C000000        mov     ecx,00000000C ;" ?"
81C088000000      add     eax,000000088 ;" ê"
8B38              mov     edi,[eax]
89BC8B18110000    mov     [ebx][ecx]*4[00001118],edi
2BC6              sub     eax,esi
49                dec     ecx
```

Figure 2.2  Example virus segment

Self-encrypting viruses use decryptor at the beginning of the file to decrypt the virus body on execution, and each generation of the file uses a different key that is generated when the virus is executed. This makes signature detection impossible as the virus body is changing on each infection. However, the problem is not hard as it seems since the decryptor itself has to be always unencrypted, and then virus researchers can extract the signature from it as long as the decryptor is long and unique enough [12]. Figure 2.3 shows the structure of a typical self-encrypting virus.

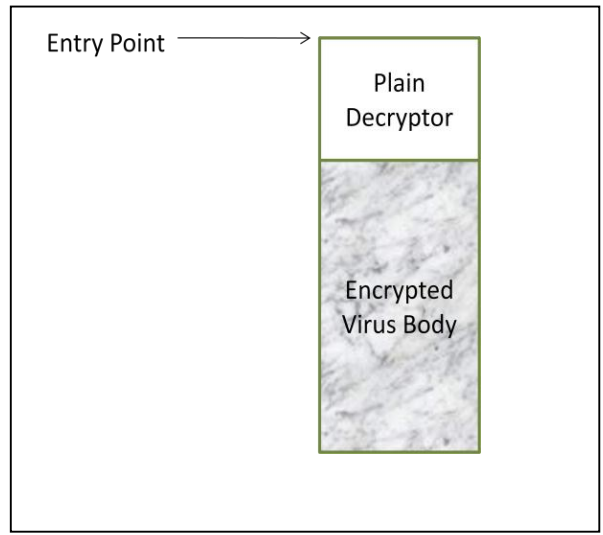

Figure 2.3  Self-encrypting Virus

Virus writers fought back with oligomorphic type of viruses, in which the virus carries some different decryptors with it, and changing the decryptor on each infection. The first known oligomorphic virus is called Whale and another famous one is Memorial which has 96 different decryptors [13]. A virus is said to be oligomorphic if it is capable of mutating its decryptor only slightly [11]. This makes antivirus researchers unable to extract a constant pattern from the decryptor to be the signature. Therefore, antivirus experts had to provide a more effective detection method, and that was the emulator. By using an emulator, the antivirus scanner can emulate code execution and after full decryption of the body, a signature can be then extracted [11].

A step further taken by attackers is the creation of polymorphic viruses. As biological viruses can be mutated in new infections, virus writers took the idea and made their virus decryptor mutate in every new infection. They attached a special module called mutation engine which responsible for mutating the decryptor to another form, yet it maintains the same behavior. Thus, polymorphic viruses can mutate their decryptors to billions of different forms, which make them virtually impossible to be detected using string signature [12].

Emulator was then the antivirus revolution. The antivirus emulator is a module that can "emulate" the execution of instructions to make the virus feel that it is on a real machine. The emulator can detect loops of decryption and after full decryption of the virus body; a signature can then be extracted and compared. Figure 2.4 shows the process of extracting signature of a polymorphic virus by the emulator.

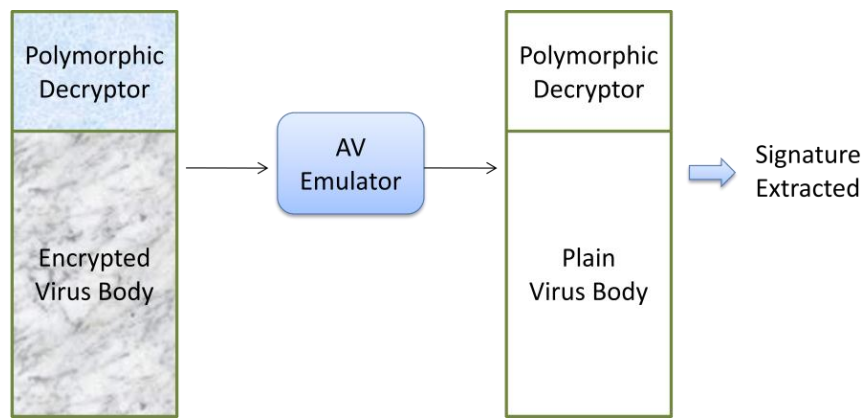

Figure 2.4  Extracting Signature from a Polymorphic Virus

An intuitively predictable step was taken after that, instead of mutating the decryptor only, virus writers mutated the entire virus body and thus encryption will not be needed any more to evade signature detection, and that was the beginning of metamorphic viruses era [14].

Metamorphic virus has a continuously changing shape for all its body, so that no constant sequence can be found in its body. Because a metamorphic virus does not need encryption to change its body, decryptor detection does not apply in this case.

In the following section we will discuss the various metamorphism techniques used by metamorphic viruses.

## 2.2 Metamorphism Techniques

Metamorphic viruses use many techniques to mutate and obfuscate their code while maintaining the same function in each generation. We will explain some of these techniques in the following subsections.

### 2.2.1 Instruction Reordering

Code obfuscation techniques used by metamorphic viruses include instruction reordering in which the virus divide its code in blocks of certain size, and then the mutation engine reorder these blocks by inserting jump instructions between the blocks while maintaining the same program result. This technique is also called code transportation and permutation [15]. Figure 2.5 shows three instruction reordering metamorphism among three generations [16].

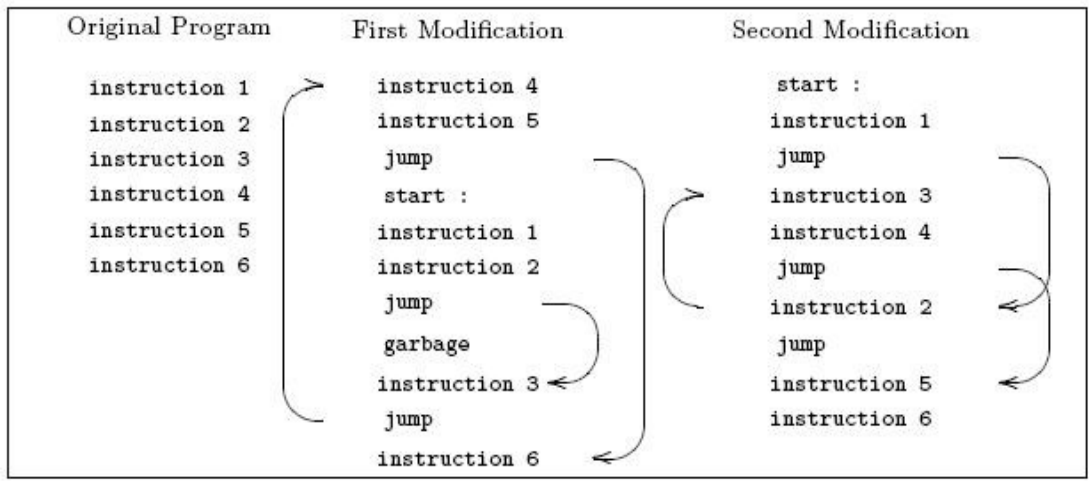

Figure 2.5  Instruction Reordering

## 2.2.2  Garbage Code Insertion

Another technique is garbage code insertion, trash insertion or dead code insertion. In this method, the mutation engine inserts useless instructions in random locations in the code, which makes the code looks very different in each generation. Examples of trash instructions are NOP which does absolutely nothing "No Operation", "mov R1, R1", "push R1" followed by "pop R1", "shl R1, 0", and many other combinations. Thus, by inserting these trash instructions in random locations in the virus, the virus has no constant body that can be detected using signature scanning [16].

## 2.2.3  Registers Swapping

Register swapping technique as it sounds is concerned with changing the registers operands of an instruction but not changing the instruction itself. An example of this type of viruses is W95/RegSwap virus. Although the resulting morphed virus looks different from the previous version, the variability is not very high and the virus can still be detected by using Half-Byte wildcard in signature string scanning [17]. Registers swapping technique is also called registers renaming or registers exchange.

### 2.2.4 Instruction Substitution

In this technique, the virus is able to replace some of its instructions with equivalent ones, while keeping the semantic of the instructions the same. This technique was used in MetaPHOR mutation engine that appeared in 2002 and in W95/Zmist virus [18].

### 2.2.5 Instructions Transposition

Transposition of instructions is the permutation of some instructions and changing their execution order. However, instructions transposition cannot be done with any group of instructions. They have to be unrelated, in other words, they are not dependent on each other.

For example, the instructions "mov eax, edx" and "add ecx, 5" have no dependency and thus can be transposed safely [19]. W95/Zmist virus that appeared in 2001 used this technique in its metamorphic engine.

Table 2.1 shows the detection difficulty of the discussed metamorphism techniques.

Table 2.1 Detection difficulty of some metamorphism techniques

| Technique | Easy | Medium | Hard |
|---|---|---|---|
| Instructions Reordering | √ | | |
| Garbage Code Insertion | | √ | |
| Registers Swapping | √ | | |
| Instruction Substitution | | | √ |
| Instructions Transposition | | | √ |

Due to these morphing techniques used by metamorphic viruses, detection of such viruses is extremely hard and different from usual detection techniques. Because once the virus analyst finds an appropriate unique pattern of bytes in the virus body, the virus changes itself to be very different from the previous generations. Therefore, other techniques must be sought in order to detect such viruses.

In the next chapter we will show some techniques used to detect metamorphic viruses, subsequently we will explain the proposed approach used in this thesis for virus detection, then we will depict how we tested our approach and show its results, after that we will conduct an analysis of the result and evaluation of the technique.

## 2.3 Summary

Computer viruses have been through four main stages since their first appearance. The first stage was the plain virus, in which the virus keeps the same shape across its generations. Self-encrypting virus is the second stage. Self-encrypting virus consists of a decryptor and an encrypted body. The body is encrypted with a different key on each infection, thus the virus body is always changing. However, a signature can be extracted if the decryptor is long enough. The third stage is the polymorphic virus. Polymorphic virus changes its decryptor and encryption key on each infection, therefore, keeps variable decryptor and body along its generations. The latest stage is the metamorphic virus, which is simply body polymorphic. That is, there is no decryptor in the virus; however, the virus applies metamorphism techniques that are applied on the decryptor in polymorphic viruses to be on all the body of the virus.

There are many metamorphism techniques used to obfuscate the virus body. Example of these techniques is instruction reordering, with which the virus reorders its instructions to change its shape and inserts some jump instructions to maintain the same sequence of execution. Garbage code insertion is another method with which the virus inserts unnecessary instructions that does not affect the behavior of the virus. Register swapping technique is used to swap some of the used registers in the virus body, thus changes the opcode of the instructions. Instruction substitution technique concerns with substituting some instructions with equivalent ones. Instruction transposition is about changes the order of execution of some independent instructions.

# CHAPTER: III
# LITERATURE SURVEY

# 3 LITERATURE SURVEY

Although many techniques have been proposed for metamorphic virus detection, few of them reached commercial products due to their computational feasibility and acceptable range of false-positives. In this chapter, we will survey some of commercially used techniques and some other experimentally proposed ones.

## 3.1 Commercially used Techniques

Detection techniques that succeeded to reach commercial products passed a long way of heavy testing since it was first proposed. Not only the success of the technique to recognize the virus was the only factor that made it usable, but also its time and space efficiency and its low false-positive error rate that it produces. The following subsections discuss some of currently used techniques for metamorphic virus detection and discuss their weak points as well.

### 3.1.1 Geometric Detection

One of commonly used techniques in commercial antivirus applications is Geometric Detection [11]. Geometric detection technique detects the changes in the infected file structure. For example, when W95/Zmist virus infects a file, it increases its virtual size of the data section to be at least 32KB larger than the physical size, so that such files can be suspicious of being infected by W95/Zmist. Another example is Bistro.B virus, which marks its infected file with value 0x51 in the high byte of the minor linker version field. However, geometric detection is considered prone to false-positive errors as some safe run-time compressed files have the same symptoms [13].

## 3.1.2 Wildcard Scanning

Another method used commercially is Wildcard scanning, which typically used for viruses that use register swapping technique mentioned in section 1.3. Figure 3.1 shows two generations of W95/Regswap [17]. The bold bytes of opcode are constants between both generations, so that wildcard scanning can be used. The following signature can be used to detect the example in Figure 3.1 [17]:

**??04000000 8B?? ?? 0C000000 81C088000000 8B?? 89???????????? 2B??**

Where "?" denotes variable half-byte.

Some of non-common opcodes between both generations have half-byte similarity, so that half byte wildcard can be combined with byte wildcard to produce more accurate detection string as the following:

**B?04000000 8B?? ?? 0C000000 81C088000000 8B?? 89???????????? 2BC?**

## 3.1.3 Stack Decryption Detection

The techniques began when variants of Zmorph virus appeared in the wild. Zmorph virus has a polymorphic decryptor at the entry point of the infected file. Once the file is executed, the polymorphic code decrypts the virus body and store the result into the stack. Moreover, after full decryption of the virus body, it transfers the control to the stack for the body to be executed. This technique was new at that time, and emulators were having no attention to the contents of the stack for identification. So in order to defeat this type of metamorphism emulators had to evolve and be able to detect stack contents. Unfortunately, examining the contents of the stack while emulation has negative effect on performance and scanning speed [17].

```
a)
BE04000000        mov       esi,000000004 ;" ?"
8BDD              mov       ebx,ebp
B90C000000        mov       ecx,00000000C ;" ?"
81C088000000      add       eax,000000088 ;" ê"
8B38              mov       edi,[eax]
89BC8B18110000    mov       [ebx][ecx]*4[00001118],edi
2BC6              sub       eax,esi
49                dec       ecx

b)
BB04000000        mov       ebx,000000004 ;" ?"
8BCD              mov       ecx,ebp
BF0C000000        mov       edi,00000000C ;" ?"
81C088000000      add       eax,000000088 ;" ê"
8B30              mov       esi,[eax]
89B4B920110000    mov       [ecx][edi]*4[00001120],esi
2BC3              sub       eax,ebx
4F                dec       edi
```

Figure 3.1  Two Regswap infection code fragments

## 3.1.4 Subroutine Depermutation

When Zperm and Ghost viruses released, they introduced another form of metamorphism. Instead of having the virus code to be executed sequentially instruction after instruction, they divided the code into sections or frames or what the authors called "islands" of code. Then the virus binds each frame with branch instruction to keep the control flow of execution the same. On each infection the viruses change its shape by permutating the sections or subroutines in another order [17]. Some viruses of this type increase their metamorphism by inserting garbage blocks of code. This type of metamorphism offers big number of different shapes for the mutated virus. Suppose if the number of sections in the virus is *n*, then the different shapes of the virus will be *n!*. For example, if the file has 10 sections or subroutines, it would have 3,628,800 different shapes. Figure 3.2 demonstrates subroutine permutation [17].

To overcome this technique, a partial emulation can be undertaken to restore the original order of the subroutines. This rebuilding process is called depermutation. Figure 3.3 shows an example of a depermutation process for a sample of permuted code.

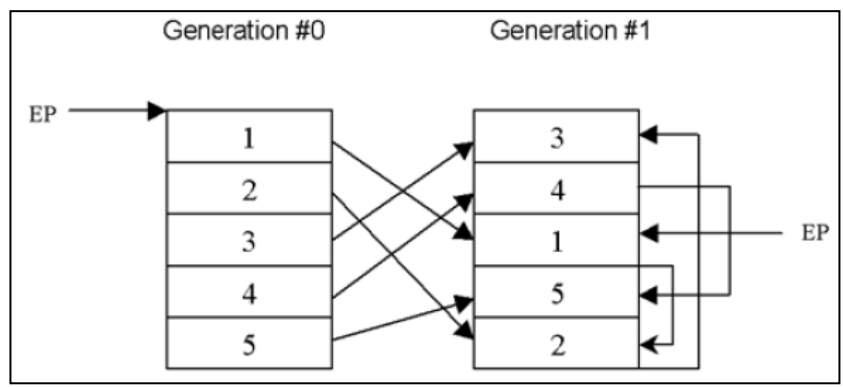

Figure 3.2  Subroutine permutation

```
Permutated code         Decoding procedure
      aaa1              1.  decode aaa1
      aaa2              2.  decode aaa2
      aaa3              3.  decode aaa3
      jmp @A            4.  change IP to @A
      bbb1              5.  decode aaa7
      bbb2              6.  decode aaa8
@B:   aaa4              7.  decode aaa9
      aaa5              8.  change IP to @B
      aaa6              9.  decode aaa4
      jmp @C            10. decode aaa5
      bbb3              11. decode aaa6
      bbb4              12. change IP to @C
@A:   aaa7              13. decode aaa10
      aaa8              14. decode aaa11
      aaa9              15. decode aaa12
      jmp @B            16. decode ret
@D:   aaa13
      aaa14
      ret
      bbb5
@C:   aaa10
      aaa11
      aaa12
      ret
```

Figure 3.3  Depermutaion process for a permuted virus.

## 3.1.5 Regular Expression and DFA

This method was discussed in details in [17], thus, this section is based mainly on [17] including figures, unless stated otherwise. In general, this method is comparatively fast compared to other techniques used for metamorphic virus detection. The method considers the input virus file as a string of alphabets or disassembly codes. These codes are compared to a database of various disassembly codes of known viruses. If a match is found, then it means that the input file is a virus, otherwise, the scanning is terminated and the file is marked as clean.

The matching of the pattern is done through the use of regular expression and DFA (Deterministic Finite Automata). In order to proceed in the section, the following terminologies have to be clear:

- **Regular expression** is a combined string of normal and special characters; this string is used to match a pattern within a target text string [20].

- **DFA** is a transition table containing states and their corresponding next states.

- **Automaton** is a predetermined sequence of operations. In this context, it corresponds to the sequence of disassembly codes.

- **Grammar** – the rules for a language. In this context, the grammar pattern relates to the set of disassembly codes that the virus uses and establishes the rule or the positive filter for detection.

The grammar pattern has information used to detect the virus, i.e. accepted instructions, and information on normalization, which is about instructions to skip or ignore (garbage instructions or negative filters). Grammar pattern uses RegEx to represent an assembly instruction.

A single disassembly code –or in other words, opcode—is an Intel IA-32 assembly instruction and an operand can be any of the following:

• <u>Exact</u> – specifies the exact operand to match.

For example:

```
PUSH EAX
```

• <u>Wildcard</u> – specifies the general type of the operand.

In case of wildcard instructions, the operand and the opcode differ. Possible wildcard values that denote registers are REG, REG8, REG16 and REG32, while the possible values for the immediate operand are IMM, IMM16 and IMM32. For memory operands, MEM, MEM16 and MEM32 are the possible values.

For example:

```
PUSH reg32
MOV reg, imm
```

`reg32` denotes that the corresponding instruction –which is PUSH in case of the first line– must be present with any 32-bit register. On the other hand, the next instruction requires that the MOV opcode is present with any register as the first operand and any immediate value as the second operand.

• <u>Variables</u> – are used to store some information on an operand and retrieved later for matching.

For example:

```
DEC reg32_varset1
PUSH reg_var1
```

For the first line, note that DEC opcode must be present with any 32-bit register, while `varset1` means to store that register type in variable 1. For the next line, the PUSH opcode must match and the operand register must also match the retrieved value of register variable 1.

The solution mainly consists of two components: the builder and the simulator.

The builder produces the automaton of the virus using the grammar pattern. Figure 3.4 shows that the pattern source is processed by DFA builder to produce automatons. In this processing, each assembly instruction is given a unique ID for later referencing and classified as garbage, accept or grammar list. Due to the fact that the pattern consists of operators, DFA builder has to deal with precedence. Therefore, for easy processing, infix expression is converted into postfix one before creating DFA patterns.

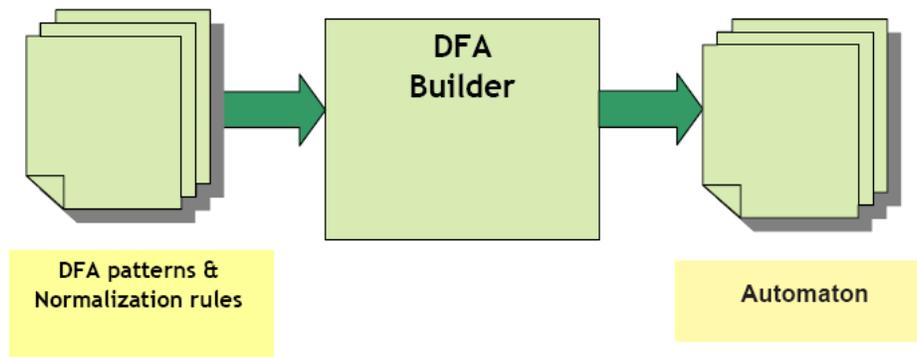

Figure 3.4 The DFA Building Process

On the other hand, the simulator performs the automaton matching and conditional test using regular expression operators during file scanning, or briefly speaking, it is responsible for scanning files for malicious content. The simulator has four sub-components: a disassembler, depermutator, normalizer and DFA simulator. Before the data is passed to DFA simulator, it has to be pre-processed by the first three sub-components. Figure 3.5 shows the simulator components.

The disassembler part converts the source from binary code to assembly code, while the depermutator attaches the subroutines of the permutated virus. The normalizer component explicitly ignores garbage instructions using the data (Garbage section) from the pattern. DFA simulation comes in the final step of the process. Using the input symbol resulting from the file being scanned and the automaton created in the building process, the DFA simulator scans the file for malicious content.

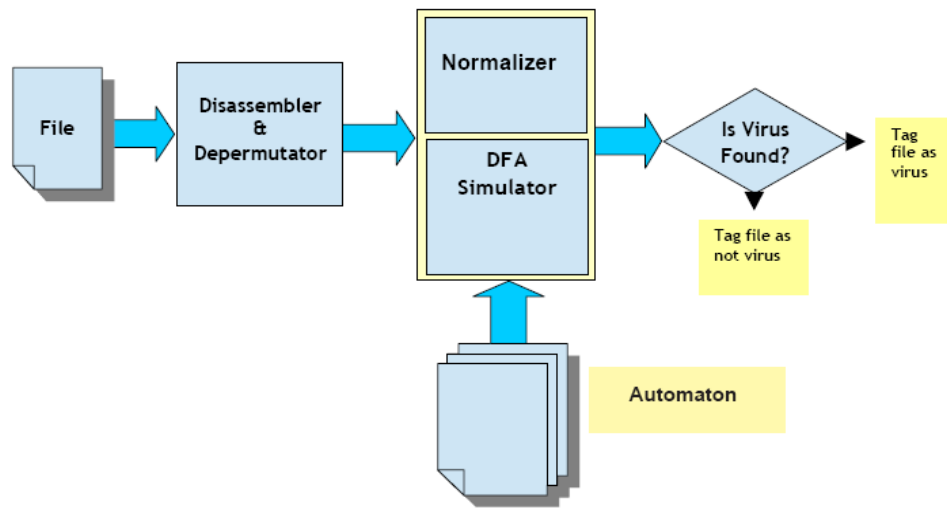

Figure 3.5 The DFA Simulation Process

The discussed solution detects almost all of the code obfuscation techniques. A virus signature for self-encrypting viruses can be creating based on the decryptor's disassembly code. Oligomorphic and polymorphic viruses can be detected by creating an automaton based on the virus' alphabets or the possible set of instructions that it can produce during infection.

Even though polymorphic viruses can produce an almost infinite number of different decryptors for each infection, these decryptors can still be split up into manageable parts, which enable the creation of a set of automatons. On most cases, these viruses can be detected generically through detection of the polymorphic engine.

Fortunately, this method also handles the detection of permutating viruses through the depermutator component, which reorders the subroutines of the permutated virus. Compared to emulators, which are known to be slow and cannot handle viruses that generate do-nothing loops, this technique basically treats the virus as a series of disassembly codes that can be matched with a database of existing virus disassembly codes. For more complicated viruses, like Zmist and Etap, this detection method works best if joined with a smart emulator.

## 3.1.6 Code Transformation Detection

This section is based mainly on [17] including figures, unless stated otherwise. Code transformation is a method of translating morphed instructions into a simplest form where common codes can be then extracted in order for the virus to be captured.

This technique was first applicable on Etap (aka Simile) virus. Etap reaches high level of metamorphism through heavy code transformation. Etap virus uses a combination of metamorphic methods such as entry point obfuscation, permutation, and heavy code mutation through shrinking and expanding techniques which is sometimes called "accordion model". To depict how highly metamorphic the virus could be, Figure 3.6 and Figure 3.7 show two generations of Etap that share the same behavior. At the first moment, the two code fragments seem very different. Nevertheless, detailed analysis of the code shows that they both construct the string "kernel32.dll" in the stack and then call "GetModuleHandle" API.

```
mov     eax,06E72656B  ;"nrek"
mov     [edx],eax
mov     eax,032336C65  ;"231e"
mov     [edx][04],eax
mov     eax,06C6C642E  ;"lld."
mov     [edx][08],eax
xor     eax,eax
mov     [edx][0C],eax
call    .00040299D
```

Figure 3.6 First generation of Etap

```
push 6c6c442e         ; mov ebp, "lld."
pop ebp
mov edx,73b36c67      ; mov edx, "231e" -> encrypted
and edx,3e7fdedd
push 4e72454b         ; mov esi, "nrek"
pop esi
push ebp              ; mov ecx, ebp
pop ecx
mov dword ptr ds:[42268c],ecx    ; mov mem+8, ecx
lea ebx,dword ptr ds:[esi]       ; mov ebx, esi
mov dword ptr ds:[422684],ebx    ; mov mem, ebx
mov dword ptr ds:[422688],edx    ; mov mem+4, edx
push 0         ; xor reg2, reg2 or mov reg2, 0
pop edx
mov dword ptr ds:[422690],edx    ; mov [mem+c], edx
mov ecx,infect1.00422684         ; mov ecx, mem
push ecx                         ; push ecx
push <&kernel32.getmodulehandlea>      ; mov edi,
offset getmodulehandle
pop edi
call dword ptr ds:[edi]; call getmodulehandle via edi
```

Figure 3.7 Second generation of Etap

To achieve this kind of high mutation, the virus code undergoes through several steps as in Figure 3.8 [21]:

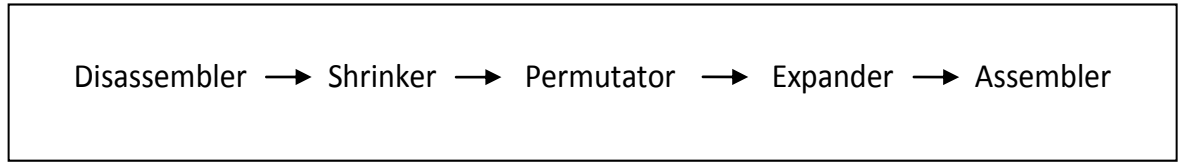

Figure 3.8 Etap virus mutation process

As explained in Figure 3.8, Etap has five main components to accomplish its metamorphism. It uses the embedded disassembler to decode each of its instructions and collect information about instruction length and used registers. The shrinker is responsible of compressing the decoded instructions by substituting one, two or three instructions with an equivalent single instruction; in addition, removing garbage codes and do-nothing loops is done at this stage. Figure 3.9, Figure 3.10 and Figure 3.11 shows sample Win32 instructions that Etap has compressed/transformed. The next step is using the permutator, in which the virus reorders its code blocks to increase the level of metamorphism. The expander simply reverses what the shrinker did. It transforms the single instructions into corresponding singles, pairs or triplet instructions. In the final step, the assembler's task is to convert the pseudo-assembly code into the real Intel IA-32 assembly instructions.

```
XOR Reg,-1          -> NOT Reg
SUB Mem,Imm         -> ADD Mem,-Imm
XOR Reg,0           -> MOV Reg,0
ADD Reg,0           -> NOP
AND Mem,0           -> MOV Mem,0
XOR Reg,Reg         -> MOV Reg,0
SUB Reg,Reg         -> MOV Reg,0
AND Reg,Reg         -> CMP Reg,0
TEST Reg,Reg        -> CMP Reg,0
LEA Reg,[Imm]       -> MOV Reg,Imm
MOV Mem,Mem         -> NOP
```

Figure 3.9 One-to-one instruction transformation.

```
PUSH Imm / POP Reg              -> MOV Reg,Imm
MOV Mem,Reg / PUSH Me           -> PUSH Reg
MOV Mem,Reg / MOV Reg2,Mem      -> MOV Reg2,Reg
ADD Reg,Imm / ADD Reg,Reg2      -> LEA Reg,[Reg+Reg2+Imm]
OP Reg,Imm / OP Reg,Imm2        -> OP Reg,(Imm OP Imm2)
LEA Reg,[Reg2+Imm] / ADD Reg,Reg3   -> LEA Reg,[Reg2+Reg3+Imm]
POP Mem / PUSH Mem              -> NOP
MOV Mem2,Mem / MOV Mem3,Mem2    -> MOV Mem3,Mem
OP Reg,xxx / MOV Reg,yyy        -> MOV Reg,yyy
NOT Reg / NEG Reg               -> ADD Reg,1
NEG Reg / ADD Reg,-1            -> NOT Reg
```

Figure 3.10 Two-to-one instruction transformation

```
MOV Mem,Reg
OP Mem,Reg2
MOV Reg,Mem         -> OP Reg,Reg2

MOV Mem,Imm
OP Mem,Reg
MOV Reg,Mem         -> OP Reg,Imm (it can't be SUB)

MOV Mem2,Mem
OP Mem2,Imm
MOV Mem,Mem2        -> OP Mem,Imm

CMP Reg,Reg
JNO/JAE/JZ/JBE/JNS/JP/JGE/JLE @xxx
!= Jcc              -> JMP @xxx

MOV Mem,Imm
CMP/TEST Reg,Mem
Jcc @xxx            -> CMP/TEST Reg,Imm
                       Jcc @xxx
                       Jcc @xxx

MOV Mem,Reg
AND/TEST Mem,Reg2
Jcc @xxx            -> TEST Reg,Reg2
                       Jcc @xxx

MOV Mem2,Mem
SUB/CMP Mem2,Reg
Jcc @xxx            -> CMP Mem,Reg
                       Jcc @xxx

MOV Mem2,Mem
AND/TEST Mem2,Imm
Jcc @xxx            -> TEST Mem,Imm
                       Jcc @xxx

MOV Mem2,Mem
SUB/CMP Mem2,Imm
Jcc @xxx            -> CMP Mem,Imm
                       Jcc @xxx

PUSH EAX
PUSH ECX
PUSH EDX            -> APICALL_BEGIN
```

Figure 3.11 Three-to-one/two/three instruction transformation

Etap virus detection has three possible solutions – simple string search, behavior checking, and code transformation. The first and second methods do not give perfect detection and produce some false positive errors. The third method is the most suitable solution for this type of metamorphism, but is also very hard to implement.

Most anti-virus engines already support string search, and it was already discussed in details earlier, so it will not be discussed here. The second method requires an emulator to trace the virus code and activate several flags when a behavior that relates to the virus is encountered.

However, because of the fact that API names cannot be resolved properly in some virus samples, this technique does not guarantee perfect detection. In addition, an emulator is required to intercept real-time instructions such as RDTSC instruction and ensures that correct values are specified so that the virus continues its execution. Otherwise, the virus simply terminates and the scanner fails to observe the virus behavior, resulting in a missed detection. Another disadvantage of this method is that it is slow – because it requires the emulation of every Intel IA-32 instruction.

On the other hand, code transformation is hard to implement. The method involves transforming the virus code back to its form prior to the expander stage. The resulting form is similar to the first generation as mentioned in Figure 3.6.

In this method, the virus code is transformed into its simplest form, as the shrinker component would do, where common instructions for virus detection are applicable. Three instructions are transformed to two or one instruction(s); two instructions are transformed to one instruction.

The code transformation module has to be heavily optimized and flexible to be able to give possible perfect detection without affecting scanning performance. Checking filters via geometric techniques like file structure analysis is also desirable. Code transformation is also useful against Zmist virus that uses techniques that are similar to those used by Etap.

Table 3.1 shows the discussed commercially used techniques and their limitations.

Table 3.1 Limitations of some detection techniques

| Technique | Limitation |
|---|---|
| **Generic Detection** | High false-positive rate |
| **Wildcard Scanning** | Limited to single metamorphism technique |
| **Stack Decryption** | Limited, slow |
| **Subroutine Depermutation** | Limited |
| **RegEx and DFA** | Needs emulator with complex viruses |
| **Code Transformation** | Limited to shrink up to only three instructions |

## 3.2 Experimental Techniques

The following subsections demonstrate some proposed techniques for metamorphic virus detection. These techniques have been proposed in academic publications. However, none of them was widely used in anti-virus commercial engines by the time of writing this thesis. The techniques did not reach commercial products for one or more reasons, either their low successful detection rate or time and space infeasibility or high false-positive rate. The subsections discuss some of these techniques and their disadvantages.

### 3.2.1 Arbitrary Length of Control Flow Graphs

In 2006 a static analysis heuristic detection method by arbitrary length of control flow graphs was presented by [22], assuming the virus does not change its control flow during propagation, but if it does, the authors proposed applying nodes alignment for detection [22]. The method showed 100% successful detection of the subject test files. However, the method was applied to only two virus classes, NGVCK and VCL32, and the number of test sample of each was 60 files, which is not enough number to show the efficiency of the method. In addition the method was not applied on other hard metamorphic viruses, such as W95/Zperm or W32/Simile. The paper also did not mention anything about how the method performs when testing benign files. As a result, percentage of false-positive errors of the method was not defined.

## 3.2.2 Zeroing Transformation

Another method is zeroing transformation method that is used to reverse the effect of some obfuscation techniques done by the mutation engine. The resulting form of the program after applying zeroing transformation on it is called zero form. Their method showed considerable decrease in the number of variants of subject programs considered in their test [23]. However, The Zeroing Transformation method does not work against expression rewriting at a low level. For example, the statements in Figure 3.12(a) are equivalent to Figure 3.12(b). However, zeroing transformation cannot recognize that [24].

## 3.2.3 Hidden Markov Model

Hidden Markov Model (HMM) is another method for metamorphic virus detection proposed by Wing Wong and Mark Stamp [25]. Hidden Markov Model is a statistical model was first presented by Leonard E. Baum and then it was used in pattern and speech recognition [26], after that it began to be used in biological sequences and DNA analysis [27]. HMM was used by Wing Wong and Mark Stamp to detect metamorphic viruses [25]. The authors were able to distinguish between NGVCK virus samples and normal files with some false-positive errors. They showed good results in detecting some metamorphic viruses that have relatively high similarity among generations. However, the authors did not show the results if a well-known hard metamorphic virus is tested such as W95/Zmist, W95/Zperm or W95/Bistro. Besides, it suffers from unacceptable rate of false positive when testing normal non-virus files against the system [28].

```
a)

OR      EAX, 81818181
OR      EAX, 42424242
SUB     EAX, C3C3C3C3
OR      EAX, 3C3C3C3C
ADD     EAX, C3C3C3C4

b)

MOV     EAX, 00000000
```

Figure 3.12  Two equivalent code fragments

## 3.2.4 Static Analyzer of Vicious Executables (SAVE)

Authors of [29] proposed a static analysis tool called SAVE which stands for "Static Analyzer of Vicious Executables". The tool uses a technique of identifying the malicious files with a sequence of API calls and not sequence of instruction as in string scanning. The method first decompresses the binary file (if it was packed), then it parses the binary file to extract the API sequence used by the file. After extracting the sequence it is compared to a database of other malicious files sequences using a similarity measure. The method also uses optimal sequence alignment algorithm to align the API sequence to the compared sequence of the virus database and tests the similarity between them. If the similarity was above certain threshold, then the test file is identified as a known malicious program. Else the file is tagged as cleared. Figure 3.13 shows the operation of SAVE tool [29]. The tool uses three distance measures to identify the sample, the cosine measure, the extended Jaccard measure and the Pearson's correlation measure.

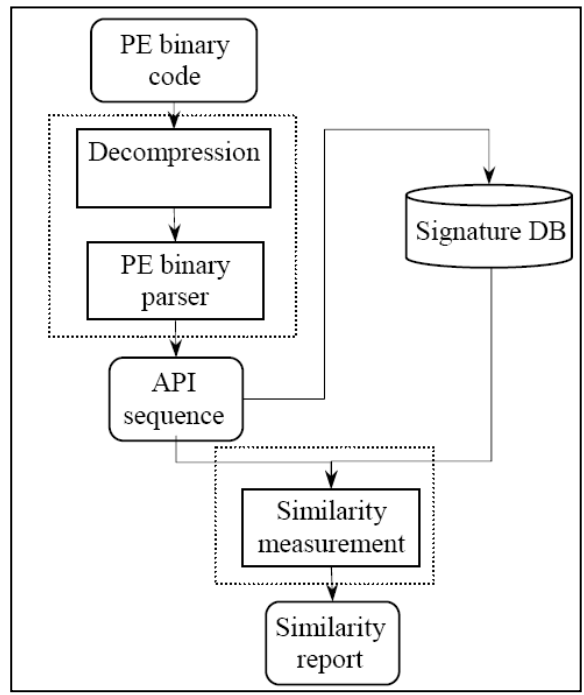

Figure 3.13 Static Analyzer for Vicious Executable (SAVE)

The authors showed that when the samples stated in the paper are modified manually, the commercial antivirus scanners fail to detect them, while SAVE tool succeeded to recognize all the samples. However, the different variants shown in the paper were generated by manual obfuscation and were not by the sample polymorphic engine. Also the samples were not of well-known hard metamorphic viruses; rather most of them were worms. In addition, the technique is mostly prone to false-positive errors, as API calling sequence can have high similarity between different viruses as they have similar behavior.

Metamorphic virus detection is still an open problem in computer virology science. There is no high performance and guaranteed method for detecting a wide range of this type of viruses [28] [30], yet some commercial techniques are doing a good job until now.

Table 3.2 shows the discussed experimental methods and their limitations.

Table 3.2 Limitations of some experimental techniques

| *Technique* | *Limitation* |
| --- | --- |
| **Arbitrary Length of Control Flow Graphs** | Easy to bypass by obfuscating control flow instructions. |
| **Zeroing Transformation** | Bypassed by low level expression rewriting. |
| **Hidden Markov Model** | Unacceptable rate of false-positive |
| **Static Analyzer of Vicious Executables (SAVE)** | Easy to bypass by obfuscating API names. |

## 3.3 Summary

Current techniques for metamorphic virus detection can be categorized into two categories. The first one is the practical or commercial techniques, while the second category is the experimental techniques. In this chapter, six commercial techniques have been discussed. The first technique is geometric detection, which concerns with the changes of the infected file in its structure and map these changes to known viruses that cause such changes. However, the method is prone to false-positive and not effective against many metamorphic viruses. Another method is wildcard scanning, which is very effective against viruses that use only register swapping technique to obfuscate their code. Stack decryption detection is another method for detection which is used only against viruses that decrypt its code in the stack. However, examining the contents of the stack adversely affects the scanning performance. Subroutine depermutation is used to reorder virus's code blocks. This method is very effective against permutated viruses such as W95/Zperm. Using regular expression with DFA is another effective method against viruses such as MetaPHOR. However, it is hard to implement and considered slow when used to detect complex viruses as it needs to be coupled with smart emulator. The last discussed method is code transformation. In this method, the virus is translated to a basic form by removing garbage code, depermutating the virus body and substituting some instructions with fewer equivalent ones.

There many experimental techniques proposed to countermeasure metamorphic viruses. Nevertheless, many of them did not reach commercial products due to their complexity or unacceptable range of false-positive errors. Arbitrary length of control flow graphs is a proposed method to detect viruses that does not change its execution flow during execution, but if it does, the method's authors proposed applying nodes alignment for detection. However, the authors did not apply the method on well-known hard metamorphic viruses; they also did not measure the false-positive rate of their method.  Another proposed method is zeroing transformation, which tries to reverse some code obfuscation techniques and transforms the code into a basic shorter form. However, the method cannot transform some complex expression rewriting. Hidden Markov model is a statistical analysis method used to identify common patterns among copies of the metamorphic virus. Nonetheless, the method suffers unacceptable rate of false-positive errors. Static analysis of vicious executables is another proposed method for virus detection. The method concerns with extracting the API calling sequence of the virus and comparing it with known sequences in the database. In the experiment of the method, variants were generated manually and not automatically by their polymorphic engine. Also the method is easy to bypass by obfuscating the API names inside the virus.

# CHAPTER: IV
# EIGENVIRUSES

# 4 EIGENVIRUSES

In this thesis, we present a detection approach based on a well-known face recognition technique called Eigenfaces [31]. Eigenfaces approach is widely and effectively used for face recognition problem. Eigenfaces approach states that every face is a linear combination of other basic set of faces called "Eigenfaces". The same person could have two different images due to change in age or light conditions or pose of face; in this case, the Eigenfaces differ in some basic faces, but not all of them. The method measures how much similarity and difference among the subject faces to decide if they can be mapped to a known face in the database or not. Figure 4.1 shows an original face and its basic Eigenfaces that construct it with some different weights [32].

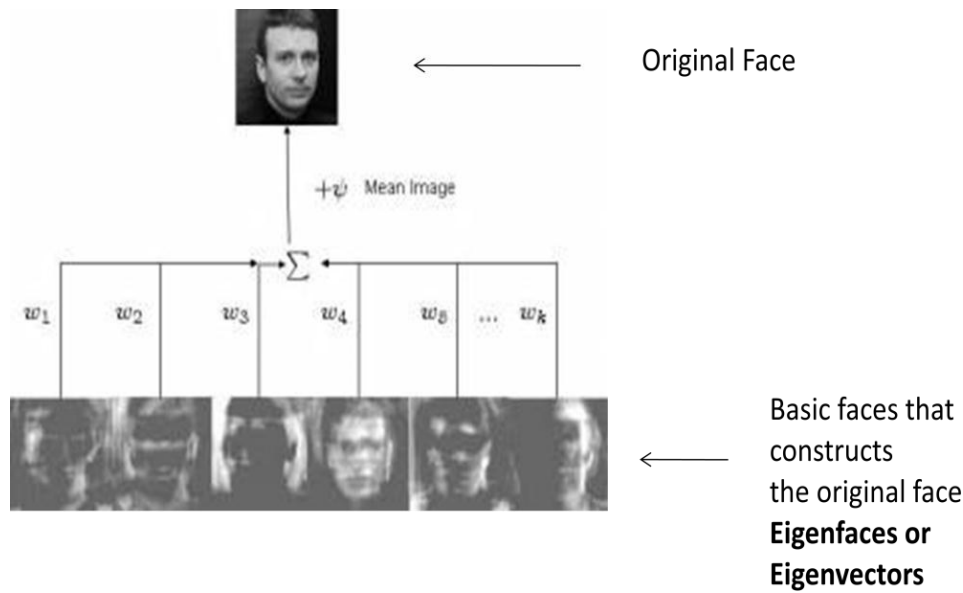

Figure 4.1  Face consists of some eigenfaces.

Our approach is using Eigenfaces technique with some modifications. As different images of the same person have some similarity among themselves, different copies of a metamorphic virus have also common similarity. Figure 4.2 generated by our system shows W95/Zperm viruses as an image (at the top), and number of its eigenviruses binaries at the bottom.

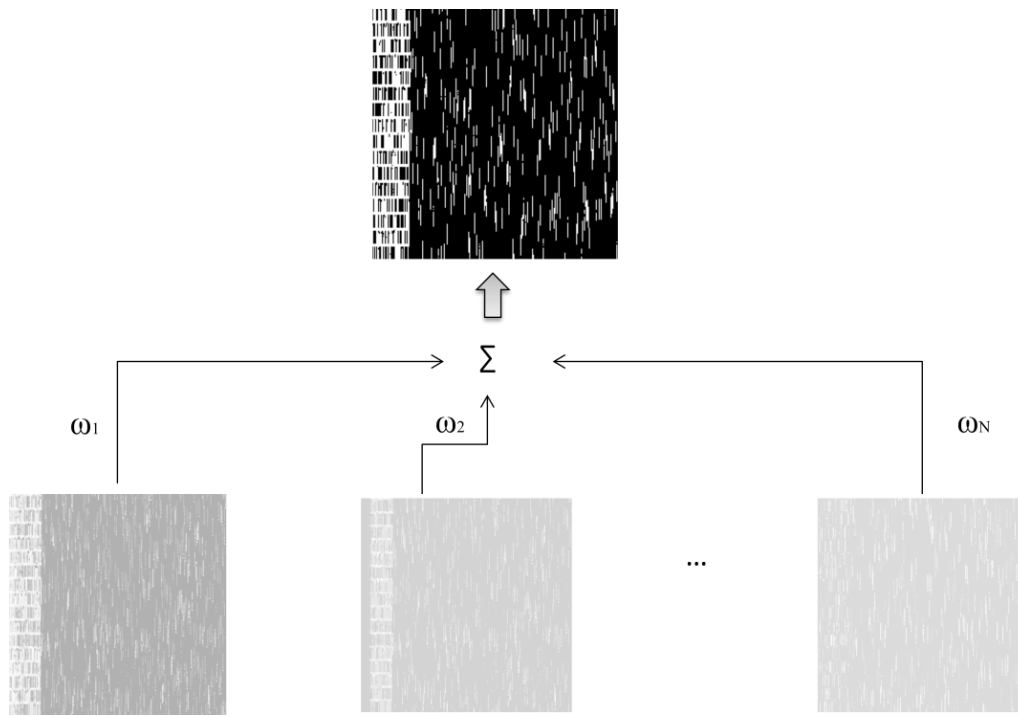

Figure 4.2  W95/Zperm virus consists of number of eigenviruses.

PCA (Principal Components Analysis) which is a statistical tool used in Eigenfaces method is used to quantify these similarities. PCA identifies the largest variances across multi-dimensional data and retains most of them. The new orthogonal vectors that span across these variations are called eigenvectors [33].

Figure 4.3 shows the eigenvectors that quantify the largest variance of a two dimensional data.

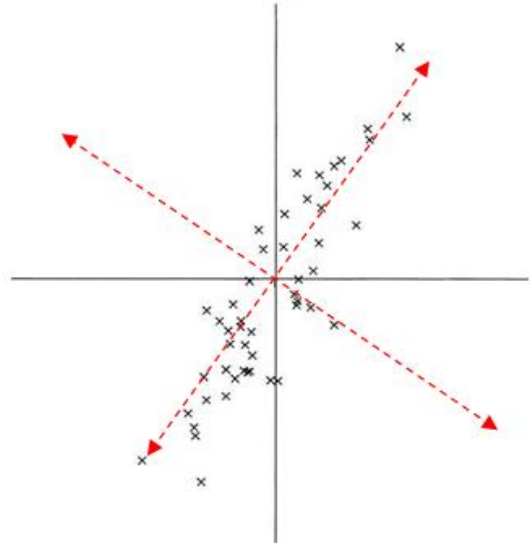

Figure 4.3 Eigenvectors of a set of 2D data

Eigenfaces approach takes advantage of principal component analysis that is used extensively in information theory. Eigenfaces approach treats the problem of face recognition as 2-D recognition problem as faces are normally upright, and ignores the geometric details of the face, which makes it relatively computationally easy and simple. The approach functions by first acquiring a set of face images, then determines the vectors or axes that span across the significant variations among the face images, those vectors are called eigenvectors, and the space defined by these vectors is called eigenspace. Since those eigenvectors when drawn give face-like images, they are called Eigenfaces.

The set of images are then projected –or in other words represented in terms of eigenvectors-- into the eigenspace or feature space, and then the system characterizes each face by weighted sum of the Eigenfaces features. Therefore, in order to determine if a new face belongs to one of the initial set or not, the new input image is projected into the eigenspace of the set of image and a distance classifier is computed between the new image's weight and each weight in the initial set. If the distance is below some threshold that was determined previously, then the image belongs to its closest class of face image, otherwise, the image does not belong to that class.

In the following parts of the thesis we will refer to "Eigenviruses" as the basic set of binaries that construct the virus which corresponds to Eigenfaces that when linearly combined constructs the face. "Training set" is the database of viruses by which our system is trained to recognize. Whereas "Test set" is the set of input viruses' replicates to be recognized. The term "replicate" refers to a morphed copy of a virus, and the term "virus" –in this context– is a general term that identifies the type of one or more replicate files such as W32/Etap, and the term "virus class" is the set of replicates that belong to a the virus. It is the mission of the technique to map an input replicate to its virus class, as Eigenfaces approach maps the input image to its face class.

The system functions by first acquiring a set of replicate files from different viruses, with more than one file from each virus. This set will be the training set of our model, then we determine the vectors or axes that span across the significant variations among the replicate files, those vectors are called eigenvectors and they construct a space called eigenspace. Since those eigenvectors when linearly combined together with certain weights they give one of the original virus replicates according to the weight, then they can be called "Eigenviruses". The set of the original replicates are then projected into that eigenspace or feature space constructed by the eigenviruses by finding the weights of each replicate. Thus, the system characterizes each virus replicate by weighted sum of eigenviruses. Then in order to determine if a new virus belongs to one of the initial set or not, the new input virus replicate is projected into the eigenspace of the set of the initial virus replicates and a distance classifier is computed between the new replicate's weight vector and each weight vector in the initial set. If the distance between the input file and the closest replicate vector in initial set is below some threshold that was determined previously, then they belongs to the same virus, otherwise the replicate does not belong to that virus.

Figure 4.4 shows the steps required to test an input file against the system.

## 4.1 Preparations

In order to apply the Eigenfaces approach on binary files, some preparations and modifications had to be made to the approach. First, it is important to remove any data in the file that is not directly relevant to the virus body. In order to do this, we removed the PE header of the file to be examined as it is not important in virus recognition in our approach, and then we extracted the malicious code of the executable file to be tested and save it into a file. By using infection flags, certain sections of PE files could only be subject to test, which reduces the complexity of the test. We also emphasize that the proposed approach focuses on recognition and classification of a malicious pattern and it is not responsible of locating and extracting the malicious pattern from a file.

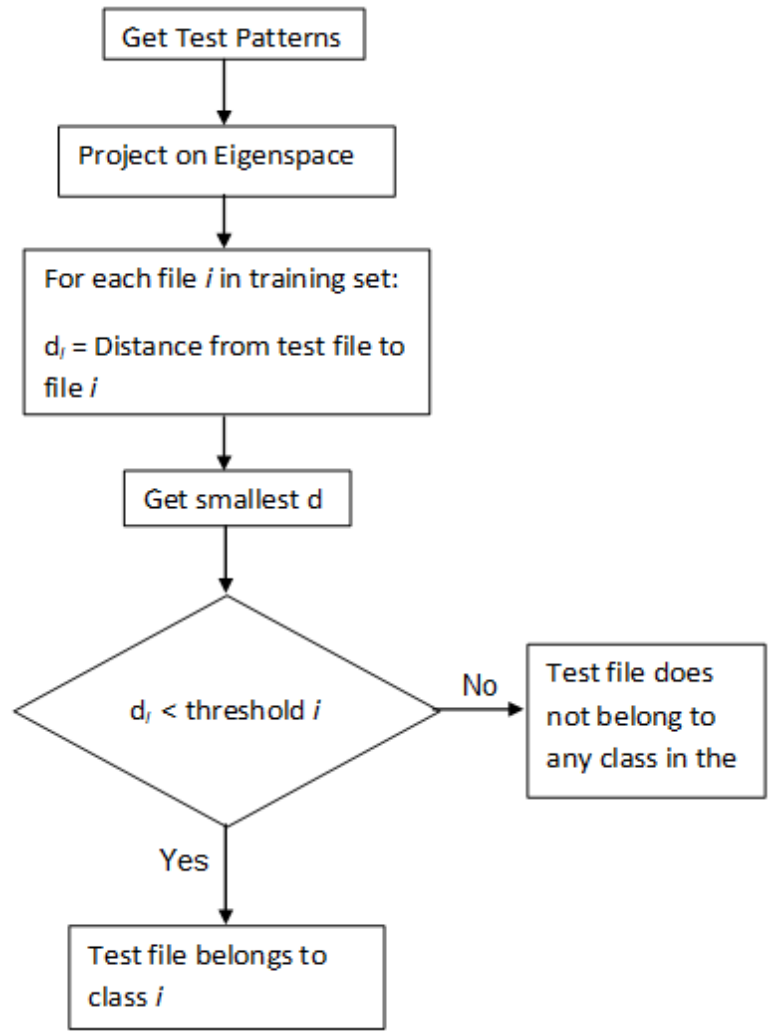

Figure 4.4 Steps to test a new input file against the system.

Because the approach requires that all inputs have to be in the same length, the input code is padded with zeros to a certain length specified when building the training set. This length is called "Eigenvirus length". The Eigenvirus length is decided based on the largest virus in the training set. Based on this length, every other input file or other virus replicate in training set that has shorter length must be padded to the eigenvirus length. If an input file is larger than the eigenvirus length, then the virus is chopped from the end to be equal to that length.

Unlike the original approach of Eigenfaces, we did not remove mean vector of the samples. Removing the mean face of face samples in Eigenfaces approach seems intuitive as all faces of different people have obvious common features, so that removing common features makes the remaining features more descriptive for the face. However, when working with binary virus files, this is not the case. Because not only different viruses can look very different, but also they can look similar to normal applications, so that subtracting the mean vector was not applied here. In addition, the original Eigenfaces approach considered one space threshold for the entire eigenspace, while we compute M space thresholds for the M virus classes in the training set. This will be further explained in the next section as well as in section 4.4.

## 4.2 Model Description

In this section, we will describe the algorithm used to project the set of virus replicates into the eigenspace, as well as how a new input replicate can be recognized as belongs to a virus class in initial set.

To construct the training set, the following steps is done:

1- Acquire an initial set of virus replicate files. M training replicate files.

2- Determine the largest file size; let us say of size N bytes. Then pad the other files with zeros to be all of size N.

3- Represent each virus replicate as a column vector $\Phi$. Therefore, $\Phi$ is an N x 1 vector.

4- Incorporate all individual virus replicate vectors into one N x M matrix A.

$$A = [\Phi_1 \Phi_2 \Phi_1, \Phi_2 \ldots \Phi_M].$$

5- Find the eigenvectors $u$ of the covariance matrix C, where $C = AA^T$. However, since C would be $N \times N$ which is computationally infeasible to get its eigenvectors for large viruses, and also C is not needed in any further computations, we should obtain eigenvectors of C without computing the value of C itself.

Suppose a matrix $L = A^T A$, where L is M x M matrix and $v_i$ is an eigenvector of L. So

$$A^T A\, v_i = \lambda_i\, v_i$$

Where $\lambda_i$ is the eigenvalue, by multiplying both sides by A it yields,

$$A\, A^T A v_i = \lambda_i\, A v_i$$

However, $C = A\, A^T$, so $A v_i$ is an eigenvector of C. As a result, if v is the set of M eigenvectors of L, then Av is the set of eigenvectors of C.

Hence $u = Av$, then we can use v to calculate the M largest eigenvectors of C

where M << N as M is the number of training virus replicates.

6- Sort the eigenvectors according to their associated eigenvalues. The higher the eigenvalue, the more important is the eigenvector in describing the features.

7- We can then choose a number of eigenvectors M` with high eigenvalues to describe the eigenspace, since not all eigenvectors represent important features of the space.

8- When projecting each virus replicate $\Phi_i$ into the eigenspace, each replicate can be represented as a linear combination of eigenvectors and weights.

$$\Phi_i = \Sigma^{M`}_{j=1} \omega_j \mu_j, \text{ where } M` <= M$$

The weights for each replicate *i* can be calculated as:

$$\omega_j = \mu^T_j \Phi_i \qquad , j = 1, 2 \ldots, M`$$

The weights of the replicate can be combined into a vector $\Omega$, where:

$$\Omega_i^T = [\omega_1, \omega_2 \ldots \omega_{M`}].$$

The previous steps were necessary to initialize the system, after that, the following steps are used to recognize a new input file:

1- Project the input file $\Phi$ $\Phi$ into the eigenspace and determine its weights.

$$\omega_j = \mu^T_j \Phi \quad , j = 1, 2 \ldots M`$$

$$\Omega^T = [\omega_1, \omega_2 \ldots \omega_{M`}].$$

2- Determine how much the input file is close to a certain virus class by measuring the Euclidean distance from its weights vector to the nearest virus replicate weight vector in the training set. This distance is called "virus class distance" $\varepsilon$.

$$\varepsilon_k^2 = \| \Omega - \Omega_k \|^2$$

$\varepsilon$ should be less than a threshold $\theta$, which is determined heuristically.

3- If we consider all the M eigenvectors to construct the eigenspace, then when a virus replicate is projected in the eigenspace, it can then be reconstructed back perfectly, as we did not ignore any of its features. However, since we chose M` eigenvectors where M` < M, accurate reconstruction of the virus replicate will not be achieved. So there will be a difference between original input replicate vector $\Phi$ and $\Phi_v = \sum_{i=1}^{M`} \omega_i u_i$, where $\Phi_v$ is the restoration of the eigenspace projected file vector.

This difference is called "virus space distance", and can be measured as:

$$\alpha^2 = \| \Phi - \Phi_v \|^2$$

Space distance measures how much the projected file lost from its features. In other words, it measures how much the eigenviruses represent the virus features. The lower the number, the fewer the loss, the more features are represented by the chosen eigenviruses. Since the chosen eigenviruses quantify the common features of all projected viruses in the space, the space distance can vary according to the virus. For each virus $i$, the space distance $\alpha$ of a newly belonging projected file should be below a threshold $\beta_i$.

There are four possibilities for the input file to be:

a- Near from a virus class and near from virus space of that class:

   In this case, the input file is recognized as belongs to that virus class.

b- Near from a virus class and far from virus space of that class:

   This happens when the input file does not belong to any class in the space, but when projected into the eigenspace, it loses many of its original features that make it looks like one of the candidate virus class.

c- Far from a virus class and near from virus space of that class:

   In this case, the input file also does not belong to any virus classes in the space. However, it shares some features with existing classes. False-negatives might occur in this case.

d- Far from virus class and far from virus space of that class:

   This case takes place when the input file does not belong to any class in the training set and does not share features with them as well.

## 4.3 Experiment

In this section, we will describe our simulation that was undertaken to evaluate the proposed approach. The next subsection describes the virus classes used in the experiment and other preparations

### 4.3.1 Samples used

We chose five viruses to run our test. They are as follows:

1- G2 Construction Kit:

G2 is a virus construction kit developed by "Dark Angel" the same author of "Phalcon/Skism Mass-Produced Code Generator" which is an earlier virus generator. G2 produces a COM and 16-bit EXE infectors. The kit has a configuration file that can be set to have the desired virus features, and then the kit produces assembly code according to the configurations. G2 can produce a different virus every time it runs, even though the values in configuration file remain unchanged. The kit mainly uses equivalent instructions substitution to achieve obfuscation. In our test we used version 1.0 which was released in January 1993 [34].

2- Next Generation Virus Creation Kit (NGVCK):

NGVCK is a virus creation kit written in Visual Basic that each time it runs it creates a virus code. Each virus created from NGVCK kit does the same function. However, every virus has almost completely different structure, which makes scanning the generated virus with the same scan string almost impossible [35][36]. NGVCK uses garbage instruction insertion, code reordering and register replacement techniques to obfuscate the generated virus code. NGVCK infects 32-bit executables and have multiple encryption methods; it also provides anti-debugging code inside the generated viruses. We used NGVCK v0.30 as it is a stable version that was released in June 2001. In the process of generating NGVCK sample files, we maintained the same configuration for all generated files.

3- <u>Zperm virus:</u>

Zperm was developed by the notorious virus writer "Z0mbie" in the year 2000. Zperm virus was one of the first 32-bit viruses for Windows platforms. The virus mainly uses permutation engine to change its instructions order constantly in each infection, including changing its permutation engine as well [37]. Zperm does not produce constant virus body anywhere as self-encrypting viruses do, instead, it permutates itself by adding and removing jump instructions and garbage instructions to produce a highly different versions of the virus. Therefore, detecting the virus cannot be done using scan string [38].

4- <u>MetaPHOR virus</u>:

MetaPHOR is a very hard metamorphic virus that was developed by "Mental Driller" in 2002. In fact, MetaPHOR was a challenge to antivirus researchers when it emerged. It is highly obfuscated and difficult to understand [18]. The virus uses various metamorphism techniques to produce a highly different new form of the virus on each infection. The virus –most of the time-- consists of a decryptor and a body. While the decryptor has a size of 4KB, the virus body has a size of more than 100KB. In our test, we used only the decryptor of the virus to test against the system as it is not encrypted and much smaller than the body, thus we can minimize the training set size. MetaPHOR is also called W32/Simile and W32/Etap.

5- Flibi worm:

Flibi is a metamorphic worm that changes its code and behavior across generations. Flibi uses new techniques for metamorphism than usual aforementioned ones. It has some analogies with molecular biology. While DNA consists of a string of nucleotides and three of the four nucleotides in the human body form a single codon, then multiple codons can be translated by tRNA to amino acids. Flibi creates a meta-language of equal size instructions of eight bits. The eight bits coding a single instruction in the meta-language are analogs of the three codons representing one amino acid and each x86 instruction parallels to amino acid. Therefore, as the same amino acid can be constructed by different codons, the same meta-language instruction can be constructed by different binary values, which then translated via a translation module written in x86 assembly into x86 instructions. When producing a new generation, the worm flips a random bit in the code, which in turn can change the meta-language to x86 instructions mapping, thus changing the behavior. There are two version of Flibi at the time of writing. Flibi.A that was released in late 2010 and it uses bit flip, byte-exchange and NOP-insertion mutations, and Flibi.B, which we used in our experiment, was released in mid-2011 and uses the same mutations operation as Flibi.A, plus horizontal gene transfer and a polymorphism technique [39][40].

## 4.3.2 Preparations

We generated two data sets for our test, a training set and a test set. The training set contains samples virus replicates of each virus; the system uses these samples to learn about each virus. Number of samples needed for each virus differs according to the virus, the more metamorphism used, the more samples needed. For constructing the training set, we needed 1, 6, 8, 15 and 2 samples of G2, NGVCK, Zperm, MetaPHOR and Flibi respectively, so our training set has 32 files. We chose eigenvirus length to be 64KiB, so that all subject files can quietly fit in that length. On the other hand, the test set contains replicates of each virus to test against the system after learning process completes. Test set contains 250 different replicates of each virus so the total number in the test set is 1250 different files.

After constructing the training set, we had 32 eigenviruses that describe the features of the training set. We chose only three eigenviruses that have the highest eigenvalues to construct the eigenspace, yet with only three eigenviruses the system showed very good results. The number of eigenviruses needed to construct descriptive space that holds the most features of viruses' classes is done heuristically.

### 4.3.3 Results

In our simulation, we constructed the training set and then we tested each file in the test set against the system, then we determined the closest file in the training set to the input virus. If there were both belong to the same virus, then it means correct detection, otherwise, the input file cannot be correctly classified. Table 4.1 shows the samples needed in the training set for each virus and the result of testing each virus replicate against the system.

Table 4.1 Results of test set against the system.

| *Virus* | *Samples Needed* | *Correctly Matched* | *Percentage* |
|---|---|---|---|
| G2 | 1 | 250 | 100% |
| NGVCK | 6 | 250 | 100% |
| Zperm | 8 | 250 | 100% |
| MetaPHOR | 15 | 250 | 100% |
| Flibi | 2 | 250 | 100% |

Space and class thresholds of each class are determined based on the result of testing the known test set. In our test, we chose the maximum values of space and class distance among the correctly matched files, so that these values correspond to the boundary of the virus class, by which we can classify an unknown input. A point worth to be noted is that, the threshold values are determined heuristically i.e. it is not necessary to choose the highest values among the correctly matched samples, but also a larger value can be chosen that mostly guarantees that all other samples from the class will lie within the class boundary. Table 4.2 shows the threshold of each of the five virus classes in our training set.

Table 4.2 Class and space thresholds for each virus.

| Virus | Class Threshold | Space Threshold |
|---|---|---|
| G2 | 50 | 3000 |
| NGVCK | 232 | 4500 |
| Zperm | 2444 | 13683 |
| MetaPHOR | 496 | 5857 |
| Flibi | 4267 | 13486 |

When constructing the eigenspace, replicates from the same class are distributed near each other according to the similarities among them. Standard deviation of a group of replicates of the same class will be a good similarity measure for files from this class. Table 4.3 shows the standard deviation of each class in the training set across each dimension. The values in Table 4.3 are rounded up to the nearest integer.

As we chose three eigenviruses to construct the eigenspace, we could then plot the viruses in a 3D space, and see how they are distributed. Figure 4.6 shows the sample files in the training set distributed in the 3D eigenspace where each axes in the space is an eigenvirus. Figure 4.5 shows the test set samples distributed in the eigenspace.

Table 4.3 Standard deviation of each virus class across 3D space.

| Virus | X | Y | Z |
|---|---|---|---|
| G2 | 0 | 0 | 0 |
| NGVCK | 128 | 57 | 236 |
| Zperm | 957 | 1896 | 2078 |
| MetaPHOR | 247 | 84 | 201 |
| Flibi | 134 | 12 | 82 |

## 4.3.4 Testing Benign Files

To measure the false positive errors in our system, in which a benign program is classified as malicious, we acquired 250 programs from Cygwin [41] utilities to test against the system. We extracted the CODE or .text sections (which represent the executable sections in most files in general) from all the files we examined and save it into a file, then the file is tested against the system. After we project the input file into the eigenspace, we determine its nearest virus class. If the projected file has distance more than the space or class threshold to its nearest class, so the file is correctly classified as does not belong to the space. Otherwise, the file is misclassified as one of virus classes and then a false-positive error produced. In our test, ten input Cygwin file was misclassified as virus. After projecting the input files into the eigenspace, 244 files had a distance more than the threshold specified for each class. This means we had 97.6% correct identifications of the sample normal files and 2.4% of false-positive errors for the subject samples.

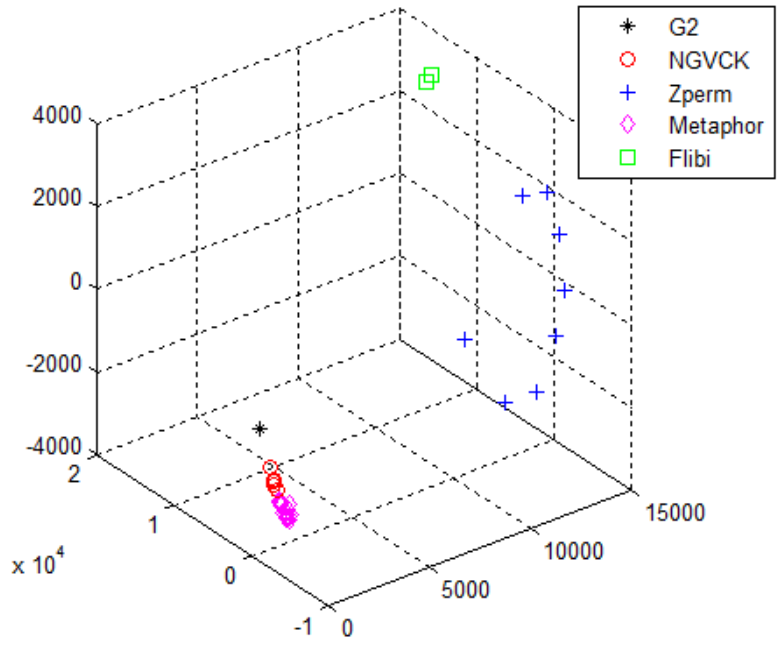

Figure 4.6 Training set virus replicates in a 3D eigenspace.

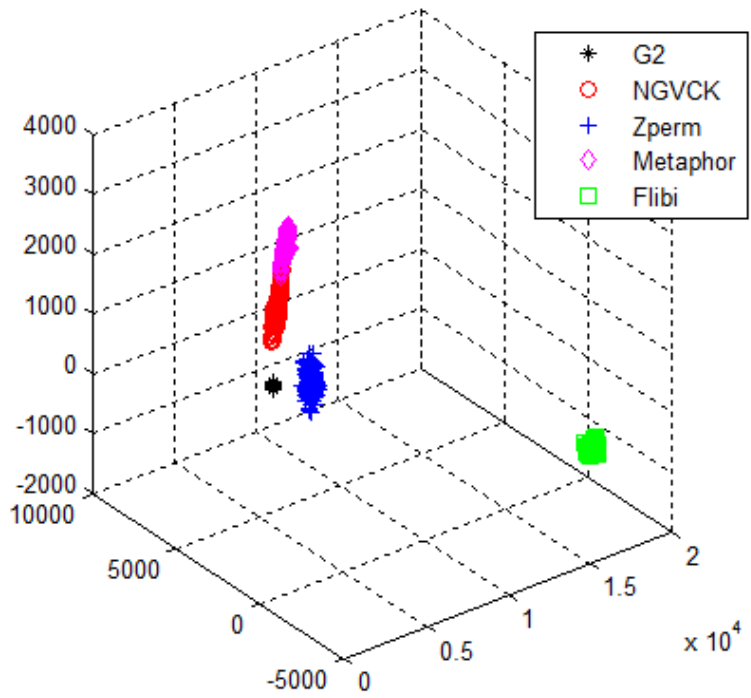

Figure 4.5 Test set virus replicates in a 3D eigenspace

## 4.4 Analysis and Evaluation

Space distance is an indicator to how much the input virus belongs to the eigenspace. The lower the number, the more feature description done by the chosen eigenviruses that construct the space. As Table 4.2 shows, G2 has the lowest class and space thresholds, which means that its samples are not scattered across the eigenspace, rather it is somehow confined in a small space and the three chosen eigenviruses could describe most of its features. Whereas Zperm has the highest space threshold due do its high variability and dissimilarities among its replicates.

Standard deviation is also a very good measure of the similarity among replicates of the same virus, as standard deviation measures the dispersion of replicates from their mean point. By examining Table 4.1 we can notice that G2 has zero standard deviation as we needed just one sample. On the other hand, Zperm has the highest standard deviation among other classes. Zperm uses code reordering extensively in a way that is not used by the other three viruses, and that is the reason why its replicates have such a high variability.

Dispersion of Zperm in the eigenspace with such comparatively high standard deviation can lead to false-positive errors with some benign files, as a projected benign file can lie anywhere in the eigenspace. The less the standard deviation of each virus class in the space, the less likely a false-positive error would occur. Normalization techniques can greatly help reduce the variability of the subject input file if it is used to preprocess the file before the test [42]. In case of Zperm, we believe that the use of a depermutator to preprocess the file would greatly reduce its standard deviation.

There is a tradeoff between database (training set) size and accurate recognition results. Since the training set represents the knowledgebase by which the system can learn about viruses, so the more training data, the more features extracted from them, the more accurate results achieved. However, more space and time complexity arise. Number of replicates for each virus needed for accurate recognition differs according to the virus. For viruses that have high similarity among their replicates, few samples are needed to construct a good model, while the opposite for hard metamorphic viruses. The size of the initial database of the system can be determined heuristically by increasing the number of replicates in each class to reach acceptable results. In addition, the system can have a continuous learning process. When the system successfully recognize a new input file as belongs to one of the classes in the database, the new file can be incorporated into the database so more features can be extracted and learned then more accurate results would be given afterwards.

To give a hint about system performance, the approach was implemented using MATLAB 7.0 R14 and ran on Windows XP SP2 and Intel Dual-core 2.60GHZ processor with 2GB RAM. It took the system about 21 seconds to scan the 1000 test files.

In order to compare the results of the system with existing detection systems, we tested a random replicate from each of the virus classes used in the experiment against three well-known antivirus products. Table 4.4 shows the result of this test [43]-[47] [1].

Table 4.4 Testing random virus replicates against commercial AVs

|  | *G2* | *NGVCK* | *Zperm* | *MetaPHOR* | *Flibi* |
|---|---|---|---|---|---|
| Symantec Norton | Detected | Heuristic | Heuristic | Detected | - |
| Kaspersky | Detected | - | Detected | Detected | - |
| ESET BitDefender | Detected | - | Detected | Detected | Heuristic |

The antivirus engines are chosen from the top of best antivirus software for the year 2009 according to AV Comparative report [48]. The versions used for Symantec, Kaspersky and ESET are 20111.1.0.186, 9.0.0.837 and 7.2 respectively.

It can be noted from the results in Table 4.4 that G2 is easy to detect so that all considered products succeeded to recognize. On the other hand, NGVCK evaded all the products except Symantec Norton that recognized the sample as a heuristic virus, i.e. the software suspects the file, but it does not recognize its name and in this case, the antivirus product will be unable to repair the infected file. In case of Zperm virus, it was successfully detected by most of the products while Symantec Norton recognized it as "Bloodhound.W32.1" which is a heuristic type of viruses and so that it cannot be repaired [49]. In case of MetaPHOR, the replicate was detected by all products, while Flibi was not recognized by any product except BitDefender that recognized it as a heuristic malware.

---

[1] DISCLAIMER: The comparison does not reflect the products' virus detection capabilities or stands as a benchmarking report. It only demonstrates how such a viral replicate can be found by the products.

## 4.5 Summary

Different images of the same person's face may seem different due to the change in pose or light direction. However, there are still common pattern among those images with which the person could be identified, and this is the main idea behind Eigenfaces approach. The same can be applied on replicates of the same metamorphic virus, thus comes the term Eigenviruses. The proposed system functions by first acquiring a set of files to be trained with. The system constructs an eigenspace where the common features of the training set represent the axes of this space. Then to test if a new file belongs to any of the viruses in the training set or not, the input file is projected into the eigenspace and its distance from each virus is calculated. If the distance between the input file and the closest virus is below certain threshold, the file is considered a morphed copy of the matched virus. Otherwise, the file does not belong to any virus in the database.

The system experiment was run on five well-known metamorphic viruses, G2, NGVCK, Zperm, MetaPHOR and Flibi. With 1, 6, 8, 15 and 2 training samples of each of them respectively. Then 250 test files of each virus were tested against the system. The result is 100% correct recognition of the test files. Also to measure false-positive errors, 250 clean files taken from Cygwin tools packages were tested against the system. Only 6 files were marked as belong to the database, this gives a false-positive rate of 2.4%.

The last section of the chapter discusses and analyses the results of the previous section. Space distance is an indicator to how much the input virus belongs to the eigenspace. The lower the number, the more feature description done by the chosen eigenviruses that construct the space. One important result is the standard deviation of each virus. The larger the standard deviation, the more metamorphic is the virus. Finally, the section shows the results when a random replicate of each used virus is tested against some antivirus engines.

# CHAPTER: VI
# CONCLUSION AND FUTURE WORK

# 5  CONCLUSION AND FUTURE WORK

This chapter concludes the work done in this thesis and intensifies the strong points and possible limitations of the proposed techniques. It also sheds some lights on possible enhancements of the system to reach better and more efficient results.

## 5.1 Conclusion

Metamorphic viruses are the hardest to detect, because of their ongoing change in structure while keeping the logical sequence the same on each infection. We developed a novel approach for metamorphic virus recognition based on a statistical machine learning technique. Our proposed approach is based on Eigenfaces technique that is generally used to solve face recognition problems. When experimented, our approach successfully recognized 100% of the test set files which consists of 1250 metamorphic virus replicates of five different hard metamorphic viruses, yet, we had 2.4% false-positive errors when 250 benign files were tested against the system.

   The proposed approach starts with a small training set that contains some replicates of each virus, and then determines the most important features among these replicates. The system represents these features by what is called Eigenviruses. Eigenviruses are vectors that span across the most important features in the sample files. By representing these sample files in terms of Eigenviruses, the recognition task is then a mere pattern recognition problem and can be solved using clustering techniques. For the five sample virus classes we chose in our test, Euclidean distance was used as a distance measure between classes. Although the used distance measure technique is very simple, it showed very good results with the chosen test set. One important advantage of the technique is that it does not depend on instructions semantic in virus' code. Therefore, common anti-debugging and anti-emulators techniques are not useful against the system.

## 5.2 Future Work

To identify the potentials of the proposed system, more viruses are needed to be tested. Due to restrictions in time and resources, only five well-known metamorphic viruses were tested. The first point as a future work is to increase the number of viruses and number of replicates of each virus in order to have a clearer picture about the capabilities of the proposed system.

With high number of virus classes used, Euclidean distance measure may not give good results. Other effective techniques can be used such as Mahalanobis distance as a distance measure or using SVM (Support Vector Machine) to cluster the groups of virus classes.

In addition, to reduce the potential errors that may occur with larger number of viruses, some malware normalization techniques can be used, such as using depermutator to help reorganize viruses that use code reordering obfuscation techniques. With such preprocessing for the input test file, we believe that the system will have great performance when used as a product. Also a detailed performance analysis should be made to accurately determine the space and time complexity of the proposed system.

# مستخلص

يعتبر الميتامورفيك فيروس هو الأخطر ضمن كل أنواع الفيروسات. بخلاف الأنواع الأخرى من فيروسات الكمبيوتر التى يمكن كشفها سواء بالطرق الثابتة مثل طريقة الكشف بالشكل الثابت للفيروس أو بالطرق المتغيرة مثل طريقة المحاكاة، يعتمد الميتامورفيك فيروس على تغيير شكله لتجنب مثل هذه الطرق، مما يجعل الميتامورفيك فيروس تحدياً حقيقياً للباحثين فى مجال أمن المعلومات. فى هذه الرسالة، سيتم دراسة الطرق التى يستخدمها الميتامورفيك فيروس لتغيير نفسه مثل إدخال أكواد عديمة الأهمية وتبديل أجزاء من الكود و اعادة ترتيب الكود وإعادة تسمية المسجلات. ثم بعد ذلك سيتم تقديم معاينة للطرق الموجودة والمستخدمه للكشف عن هذا النوع من الفيروسات وفيها سنقدم كلاً من الطرق المستخدمة فى برامج مضادات الفيروسات التجارية وتلك المقدمة فى أبحاث علمية.

ثم بعد ذلك نقدم طريقة جديدة للكشف عن فيروسات الميتامورفيك معتمدةً على طريقة التعليم بغير مُعلم المستخدمة فى الذكاء الصناعى ومستوحاة من احدى طرق التعرف على الوجوه، ثم نقوم بتحليل أداء الطريقة المقترحة ونبين نتائج تجريبية لتلك الطريقة مقارنةً مع نتائج برامج معروفة من مضادات الفيروسات. وأخيراً سنناقش التحسينات المستقبلية و المحتملة للطريقة المقترحة للتعرف على أعداد و أنواع أكثر من الفيروسات.

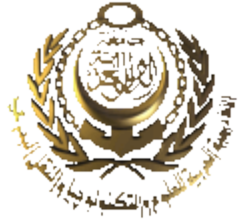

الأكاديمية العربية للعلوم و التكنولوجيا و النقل البحرى

كلية الهندسة و التكنولوجيا

قسم هندسة الحاسب الآلى

# نحو التعرف على الميتامورفيك فيروس باستخدام آيجن- فيروس

رسالة مقدمة كإستيفاء جزئى لمتطلبات درجة

**الماجستير فى هندسة الحاسب**

مقدمة من

**مصطفى السيد محمد تكرونى صالح**

بكالوريوس هندسة الحاسب بالأكاديمية العربية للعلوم و التكنولوجيا و النقل البحرى

المشرفون

| أ.د. عبد الباعث محمد محمد | د. أحمد عبد النبى |
|---|---|
| أستاذ بكلية الهندسة و التكنولوجيا | أستاذ مساعد، مدينة مبارك للأبحاث العلمية و |
| قسم هندسة الحاسب | التطبيقات التكنولوجية، معهد بحوث المعلوماتية، |
| الأكاديمية العربية للعلوم و التكنولوجيا | رئيس قسم الشبكات و التشغيل الموزع |

يوليو 2011